\documentstyle[12pt]{article}

\input{epsf.tex}

\newcommand{\E}{{\cal{E}}}

\renewcommand{\a}{\alpha}

\newcommand{\be}{\begin{equation}}
\newcommand{\ee}{\end{equation}}
\newcommand{\bea}{\begin{eqnarray}}
\newcommand{\eea}{\end{eqnarray}}

\begin{document}
\title{\bf On the discrepancy between \\ two approaches to the
equilibrium \\ problem for spinning particles}
\author{V.~S.~Manko$^\dag$ and E.~Ruiz$^\ddag$}
\date{}
\maketitle

\vspace{-0.5cm}

\noindent $^\dag$Departamento de F\'\i sica, Centro de
Investigaci\'on y de Estudios Avanzados del IPN, A.P. 14-740,
07000 M\'exico D.F., Mexico

\noindent $^\ddag$Area de F\'\i sica Te\'orica, Universidad de
Salamanca, 37008 Salamanca, Spain

\vspace{1.5cm}

\begin{abstract}{We show that the absence of equilibrium
states of two uncharged spinning particles located on the symmetry
axis, revealed in an approximate approach recently employed by
Bonnor, can be explained by a non--general character of his
approximation scheme which lacks an important arbitrary parameter
representing a strut. The absence of this parameter introduces an
artificial restriction on the particles' angular momenta, making
it impossible to find a physical solution to the balance
equations.}
\end{abstract}

\newpage

\noindent {\bf 1.} Construction of exact solutions permitting
equilibrium configurations of aligned spinning particles can be
considered as one of the most fascinating applications of the
modern solution generating techniques. Among different equilibrium
problems the superposition of two Kerr particles described by the
famous double--Kerr solution of Kramer and Neugebauer [1] is of
special interest since the corresponding balance conditions
determining an equilibrium of spinning particles can be solved
analytically in the general extended case~[2], the masses and
angular momenta of the balancing constituents verifying a very
simple relation derived in~[3]:
\be
\pm(M+s)^2+s(a_1+a_2)+J=0, \ee where $s$ is the coordinate
distance between the particles; $M$ and $J$ are the total mass and
total angular momentum of the two Kerr constituents, respectively,
related to the individual masses $m_1$, $m_2$ and angular momenta
per unit mass $a_1$, $a_2$ by the formulae
\be
M=m_1+m_2, \quad J=m_1a_1+m_2a_2. \ee

We remind that the complex Ernst potential \cite{E} of the
extended double--Kerr solution has the form \cite{MR1} \bea
\E&=&\frac{\Lambda-\Gamma}{\Lambda+\Gamma}, \nonumber\\
\Lambda&=&\sum\limits_{1\le i<j\le4}\lambda_{ij}r_ir_j, \quad
\Gamma=\sum\limits_{i=0}^4\nu_ir_i, \nonumber\\
\lambda_{ij}&=&(-1)^{i+j}(\a_i-\a_j)(\a_{i'}-\a_{j'})X_iX_j,
\nonumber\\ &&(i',j'\ne i,j; i'<j'), \nonumber\\
\nu_i&=&(-1)^i(\a_{i'}-\a_{j'})(\a_{i'}-\a_{k'})
(\a_{j'}-\a_{k'})X_i, \nonumber\\ &&(i',j',k'\ne i;i'<j'<k'),
\nonumber\\ X_i&=&\frac{(\a_i-\bar\beta_1)(\a_i-\bar\beta_2)}
{(\a_i-\beta_1)(\a_i-\beta_2)}, \nonumber\\
r_i&=&\sqrt{\rho^2+(z-\a_i)^2}, \eea where the parameters $\a_1$,
$\a_2$, $\a_3$, $\a_4$ can assume arbitrary real values or occur
in complex conjugate pairs $\bar\a_2=\a_1$ and/or $\bar\a_4=\a_3$
(a bar means complex conjugation), $\beta_1$ and $\beta_2$ are
arbitrary complex constants, and $(\rho,z)$ are cylindrical
Weyl--Papapetrou coordinates.

The original Kramer--Neugebauer solution \cite{KN} is contained in
formulae (3) as the purely black--hole case, i.e., when all $\a$s
are real quantities.

The total number of arbitrary real parameters involved in (3) is
{\it eight}; however, it reduces to {\it seven} after imposing the
asymptotic flatness condition
\be
{\rm Im}\Bigl\{\Bigl(\sum\limits_{i=1}^4\nu_i\Bigr)/
\Bigl(\sum\limits_{1\le i<j\le 4}\lambda_{ij}\Bigr)\Bigr\}=0. \ee
Furthermore, taking into account that the mass--dipole moment of
the solution can be always made equal to zero by an appropriate
shift along the symmetry $z$--axis, we obtain an asymptotically
flat {\it six}--parameter sub--family of the double--Kerr solution
which may describe two Kerr particles and a supporting strut
between them (the segment $\a_3<z<\a_2$ of the $z$--axis in
Fig.~1; note that each of the segments $\a_2<z<\a_1$,
$\a_3<z<\a_2$, $\a_4<z<\a_3$ is characterized by two arbitrary
real parameters).

\begin{figure}[htb]
\centerline{\epsfysize=80mm\epsffile{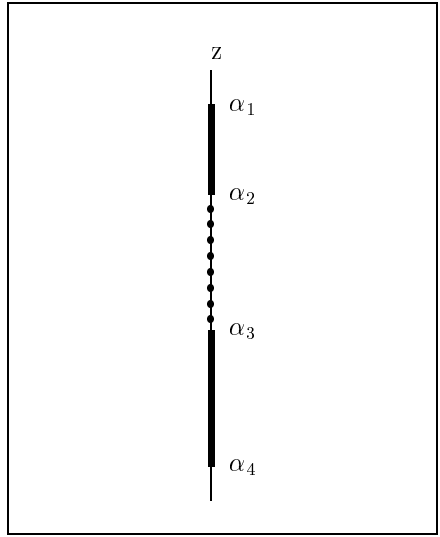}} \caption{Location
of two subextreme constituents on the symmetry axis. The
intermediate segment $\alpha_3<z<\alpha_2$ represents a strut
which can be removed by solving the balance equations.}
\end{figure}

The strut can be removed by requiring regularity of the
$\a_3<z<\a_2$ part of the axis; this implies \bea \gamma(\rho=0,
\a_3<z<\a_2)=0, \nonumber\\ \omega(\rho=0, \a_3<z<\a_2)=0, \eea
where $\gamma$ and $\omega$ are the metric coefficients in the
line element \be
ds^2=f^{-1}[e^{2\gamma}(d\rho^2+dz^2)+\rho^2d\varphi^2]
-f(dt-\omega d\varphi)^2, \ee and can be constructed from $\E$
(the explicit form of $\gamma$ and $\omega$ corresponding to (3)
can be found in \cite{MRS}; $f$ is the real part of $\E$).

Equations (5) are two balance conditions whose resolution
eventually leads to a {\it four}--parameter general class of
solutions for two Kerr particles in equilibrium (without a strut)
due to a balance of the gravitational attraction and spin--spin
repulsion forces \cite{MR1}. These four surviving arbitrary real
parameters can be associated with the individual masses and
angular momenta of the Kerr particles, and the quadratic equation
(1) defines at which coordinate distance $s$ the equilibrium takes
place for a specific choice of $m_1$, $m_2$, $a_1$, $a_2$
(evidently only positive values of $s$ are physically
admissible\footnote{It should be emphasized that Eq.~(1), derived
after solving the general balance problem, is not one of the
balance conditions (these apparently involve more parameters) but
may be considered as a corollary of the balance equations.}).

Although two Kerr black holes with positive masses can never be in
equilibrium due to a balance of the gravitational attraction and
spin--spin repulsion forces [2], equilibrium states between a
subextreme and a hyperextreme constituents, or between a pair of
hyperextreme constituents with positive masses are possible (see
Refs.~[5, 6] for concrete examples). Mention here that various
equilibrium configurations of spinning Curzon particles were
obtained long ago [7, 8].

\bigskip

\noindent{\bf 2.} Recently, Bonnor [9] tried to solve the balance
problem for two spinning particles with the aid of an
approximation method. According to~[9], two particles will be in
equilibrium if the following conditions are satisfied: \bea
s^2&=&6a_1a_2, \\ a_2&=&-a_1 \eea [in the same notations as used
in (1), (2)]. It is trivially seen that (7) and (8) yield for the
coordinate distance $s$ two {\it pure imaginary} values,
\be
s=\pm\sqrt{6}ia_1, \ee which means that according to (7) and (8)
there is no equilibrium state at all. Moreover, we have
demonstrated [3] that equilibrium configurations of two Kerr
particles from [5] possessing positive Komar masses do not satisfy
approximately even one of the equations (7) or (8), which
suggested that Bonnor's approach should be rectified.

Bonnor's Comment \cite{com} to our criticism [3] contains an
attempt to defend the physical content of the balance conditions
(7), (8), that does not look appropriate in view of Eq.~(9). The
arguments used in [10] are essentially as follows: both conditions
(7) and (8) had previously appeared in the literature in
connection with specific exact solutions, so this must lend them
credit independently of the context in which they were obtained; a
precise comparison of the extended double--Kerr solution and the
approximate formulae is not possible in view of the
``unsurveyable'' form of the former. In Ref.~[10] there was no
attempt to answer the key question of why the equilibrium states
of two spinning particles, obtainable using exact solutions, do
not emerge in the approximation scheme.

In what follows we are going to point out ($i$) why the
approximation scheme of \cite{bon} could only lead to the
physically unacceptable relations (7), (8), and ($ii$) that the
arguments employed in \cite{com} are in fact misleading.

\medskip

\noindent{\bf (i)} We identify the origin of the failure in
finding the equilibrium states within the framework of Bonnor's
approximation procedure (assuming that it is mathematically
correct) with the absence in it of an important additional
arbitrary parameter representing the torsion singularity, i.e.,
the angular momentum of the part of the symmetry axis separating
the particles.

Indeed, the approximation of Ref.~[9] uses only {\it five} real
constants to describe two spinning particles and a strut between
them, ignoring that the strut, like each particle, is
characterized by two parameters, mass and angular momentum (as we
have already shown, this case involves {\it six} constants in the
double--Kerr solution). The constants $m_1$, $m_2$, $a_1$, $a_2$
and $s$ are introduced by Bonnor via the functions $f^{(1)}$ and
$\omega^{(1)}$ in the representations of the metric coefficients
$f$ and $\omega$: \bea f&=&\sum\limits_{i=0}^2f^{(i)}, \quad
\omega=\sum\limits_{i=0}^2\omega^{(i)}, \nonumber\\ f^{(0)}&=&1,
\quad \omega^{(0)}=0, \nonumber\\ f^{(1)}&=&-\frac{2m_1}{r_1}
-\frac{2m_2}{r_2}, \nonumber\\
\omega^{(1)}&=&\frac{2m_1a_1\rho^2}{r_1^3}
+\frac{2m_2a_2\rho^2}{r_2^3}, \nonumber\\
r_1&=&\sqrt{\rho^2+(z-(s/2))^2}, \nonumber\\
r_2&=&\sqrt{\rho^2+(z+(s/2))^2} \eea (see [9] for the explicit
form of $f^{(2)}$ and $\omega^{(2)}$). We mention that two
additional constants which arise during the calculation of the
metric coefficient $\gamma$ and the function $\omega^{(2)}$ are
here needed for preserving the asymptotic flatness of this
approximate solution.

Consequently, after imposing two balance conditions similar to (5)
on the above five parameters, Bonnor ends up with only {\it three}
arbitrary parameters, thus necessarily introducing the dependence
of the angular momentum per unit mass $a_2$ on $a_1$ via (8)
(recall that $m_1$, $m_2$, $a_1$, $a_2$ are arbitrary independent
constants in the exact approach to equilibrium of spinning
particles).

The non--general nature of the above approximation scheme is
evident since the above unphysical particular branch of
``equilibrium states'' is also contained in our general formulae
describing two balancing Kerr particles. Indeed, after choosing
the `+' sign on the left--hand side of equation (1) and setting
\be
a_2=-a_1, \quad m_2=-m_1=-3a_1, \ee one immediately arrives at
Eq.~(9).

Inclusion of the missing parameter into the approximation
procedure will most probably allow one to achieve correspondence
with the known equilibrium states of two Kerr particles possessing
positive masses since the resulting solution of the approximate
balance problem will have already {\it four} arbitrarily
prescribed parameters of both particles $m_1$, $m_2$, $a_1$,
$a_2$, precisely as in our general exact solution of the
double--Kerr equilibrium problem \cite{MR1,MR2}.

\medskip

\noindent{\bf (ii)} We find it instructive to clarify some points
concerning the two exact solutions mentioned in \cite{com} (we do
not think that the remark on an ``unsurveyable'' form of the
double--Kerr solution needs a comment). First, we explain why the
known exact PIW solutions \cite{Per,IW} do not lend physical
support to the balance condition (8) in the context of pure vacuum
spacetimes. The PIW solutions describe very special electrovac
stationary spacetimes where {\it gravity} is balanced by an {\it
electric} force. Due to the rotation of sources, there also arise
two more forces, the {\it magnetic} and {\it spin--spin} ones, and
Eq.~(8) represents the necessary requirement of balance of the
latter two forces. Hence, in the balancing PIW solutions, gravity
is not affected at all by the spin--spin force (!), and the
parallel made in \cite{com} between the PIW electrovac solutions
and the approximate pure vacuum problem, where exclusively the
gravitational and spin--spin forces could balance each other, is
improper.

Another exact result incorrectly interpreted in \cite{com} is
related to a particular binary system of two identical
counter--rotating charged particles first considered in \cite{BM}.
Recall that, in \cite{BM}, condition (8) was introduced into an
exact electrovac solution by {\it construction}, and it led to a
rigorous result that, in the pure vacuum limit, an equilibrium of
two counter--rotating identical Kerr particles was {\it
impossible}. The approach of Ref.~\cite{bon} is just the opposite:
one seeks a balance of two arbitrary spinning particles and
arrives at Eq.~(8) as a {\it necessary} condition for equilibrium,
in evident contradiction to the above exact result \cite{BM} on
the non--existence of a balance of two particles with equal masses
and opposite angular momenta.\footnote{The non--existence of
balance in this case immediately follows from our relation (1):
setting $a_2=-a_1$ and $J=0$, one obtains that the coordinate
distance $s$ is equal to {\it minus} total mass $M$ of the
system.} Therefore, instead of supporting Eq.~(8), Ref.~\cite{BM}
only identifies the latter as a condition which is {\it unlikely}
to appear in the approximation scheme of \cite{bon}. It is worth
pointing out that Bonnor's suggestion to generalize the solution
of Ref.~\cite{BM} to the case of non--identical particles in order
to compare the resulting balance conditions with Eq.~(8) is
superfluous because, firstly, the general double--Kerr--Newman
solution is already known (it is the $N=2$ specialization of the
extended multi--soliton electrovacuum metric \cite{RMM}), and,
secondly, this solution in the absence of the electromagnetic
field reduces to the double--Kerr spacetime for which the balance
problem has already been solved, and the existing equilibrium
states with positive masses contradict the approximate results of
[9].

The balance problem for two uncharged spinning particles is thus a
serious test which has already been passed by exact solutions in
several elegant ways; the approximation method of [9], to become
successful, has yet to be rectified along the lines discussed
here.

\bigskip

{This work has been partially supported by Project 34222-E from
CONACYT of Mexico, and by Project BFM2000--1322 from MCYT of
Spain.}

\end{document}